\documentclass[11pt]{elsarticle} 

\usepackage{amsmath}
\usepackage{amssymb}
\usepackage{amsfonts}
\usepackage{caption}
\usepackage[hyphens]{url} 
\usepackage{graphicx} 
\usepackage{natbib}  
\usepackage{booktabs}
\usepackage{paralist}
\usepackage{bm}
\usepackage{pgfplots}
\usepackage{mathrsfs}
\usepackage{longtable}
\usepackage{rotating}
\usepackage{multirow}
\usepackage{subfigure}
\usepackage{enumitem}
\usepackage{adjustbox}
\usepackage{diagbox}

\usepackage{algorithm}
\usepackage{algorithmic}

\usepackage{newfloat}
\usepackage{listings}
\newfloat{listing}{tb}{lst}{}
\floatname{listing}{Listing}

\usepackage{xcolor}

\usepackage{chemformula}
\usepackage[margin=2.5cm]{geometry}
\usepackage{array}      % for >{...} column modifiers
\usepackage{tabularx}   % for tabularx environment
\usepackage{booktabs}   % for \toprule, \midrule, \bottomrule
\usepackage{makecell}   % for \makecell
\usepackage{threeparttable}

\begin{document}
%\nolinenumbers

\begin{frontmatter}
\title{\emph{Quantum Neural Physics:} Solving Partial Differential Equations on Quantum Simulators using Quantum Convolutional Neural Networks} 

\author[RGA,AMCG]{Jucai Zhai\corref{cor1}}
\cortext[cor1]{Corresponding author}
\ead{jucai.zhai25@imperial.ac.uk}
\author[QUAIL]{Muhammad Abdullah}
\author[AMCG]{Boyang Chen}
\author[QUAIL]{Fazal Chaudry}
\author[AMCG]{Paul N. Smith}
\author[AMCG,IX]{Claire E. Heaney}
\author[RGA]{Yanghua Wang}
\author[AMCG]{Jiansheng Xiang}
\author[AMCG,IX,DAL]{Christopher C. Pain}

\address[RGA]{{Resource Geophysics Academy}, {Imperial College London}, {{London}, {SW7 2AZ} {UK}}}
\address[AMCG]{Applied Modelling and Computation Group, Department of Earth Science and Engineering, Imperial College London, London, SW7 2AZ UK}
\address[QUAIL] {Quantum AI Leadership (QUAIL) Ltd, London, WC2H 9JQ UK}
\address[IX]{Centre for AI-Physics Modelling, Imperial-X, White City Campus, Imperial College London, London, W12 7SL UK}
\address[DAL]{Data Assimilation Lab., Data Science Institute, Imperial College London, London, SW7 2AZ UK}

%jucai.zhai25@imperial.ac.uk, j.xiang@imperial.ac.uk, c.pain@imperial.ac.uk

\begin{abstract}
Neural Physics recasts local discretisations of partial differential equations (PDEs) as fixed convolutional operators, providing a physics-preserving alternative to data-driven surrogate modelling in scientific machine learning. However, existing realizations remain largely confined to classical AI hardware and do not directly connect to quantum structured operator design. To bridge this gap, we introduce a \emph{Quantum Neural Physics} framework and develop a Hybrid Quantum-Classical CNN Multigrid Solver (HQC-CNNMG). The proposed method maps analytically prescribed stencil operators to local quantum convolutional primitives and embeds them within a classical multilevel W-cycle architecture, combining the operator-centric view of scientific ML with the numerical rigor of multigrid solvers. Using amplitude encoding together with the Linear Combination of Unitaries (LCU) and the Quantum Fourier Transform (QFT), the resulting local quantum operators admit logarithmic-depth implementation, with circuit depth scaling as $\mathcal{O}(\log K)$ for an encoded block of size $K$ under the idealized parallel circuit model considered here. Numerical experiments on Poisson, transient diffusion, convection--diffusion, and incompressible Navier--Stokes problems demonstrate numerical consistency, stable multilevel behaviour, and workflow-level feasibility on noiseless simulators. Comparisons with representative quantum linear solver paradigms further show that the main strength of HQC-CNNMG lies in its balanced trade-off among local circuit depth, numerical robustness, and compatibility with PDE structure, rather than in fully quantum global inversion.
\end{abstract}


\end{frontmatter}

\section{Introduction}
\label{sec:intro}

Large-scale scientific computing is central to modern physics, fluid dynamics, and engineering design, where the numerical solution of partial differential equations (PDEs) remains a fundamental task. Classical discretisation methods, including the Finite Difference Method (FDM) and the Finite Element Method (FEM), provide rigorous and well-established frameworks for PDE simulation, but their deployment at extreme scale is increasingly constrained by computational cost, memory bandwidth, and energy efficiency on classical hardware~\cite{Numerical_approx_PDE_Quarteroni_2008,memory_wall_wulf_1995,memory_wall_dongarra_2011}.

At the same time, Artificial Intelligence (AI), especially Deep Neural Networks (DNNs), has opened new directions for scientific computing. Data-driven surrogate models~\cite{ML_accelerate_CFD_Dmitrii_2021,pfaff2021meshgraphnet,lu2021deeponet} can accelerate prediction by learning mappings between system parameters, intermediate states, and solution fields. However, because these models depend on training data, their reliability and generalisation may degrade outside the training distribution, particularly for strongly nonlinear regimes, high-Reynolds-number flows, or problems with singular structures~\cite{brunton2020_ML_limited_CFD,KARNIADAKIS2021PIML}. This has motivated growing interest in approaches that combine the efficiency of modern AI software stacks with the rigour of classical numerical discretisations.

One such direction is the reinterpretation of local PDE discretisations as fixed convolutional operators within Convolutional Neural Networks (CNNs)~\cite{Zhao2020,WANG2022tensorflowNP,phillips2023solving,phillips2023convfem}. In this perspective, often referred to as \emph{Neural Physics}~\cite{CHEN2026Neuralphysics}, analytically prescribed stencil coefficients play the role of convolutional kernel weights. Unlike data-driven surrogate models, these operators are not learned from data, but are determined directly by the governing equations and the underlying numerical scheme~\cite{chen2024solving,chen2025solving,nadimy2025solving,naderi2024discrete,yang2026maximising,li2025implementing}. As a result, Neural Physics provides a training-free, matrix-free, and hardware-compatible framework that connects classical numerical analysis with AI-oriented computational backends such as PyTorch, TensorFlow, JAX, GPUs, and TPUs~\cite{WANG2022tensorflowNP,Bezgin2023,Zhao2020}.

Despite these advances, current Neural Physics formulations remain essentially classical. In parallel, quantum computing has emerged as a promising computational paradigm for certain structured linear algebra and operator-based tasks~\cite{montanaro2016quantum_review,childs2021QFDM}. Recent studies have explored quantum and quantum-machine-learning approaches for fluid simulation, including benchmarks of Quantum Tensor Networks, Hydrodynamic Schr\"odinger Equation formulations, and Physics-Informed Neural Networks for the Burgers equation~\cite{kerni2026benchmarkingquantumclassicalalgorithms_1refer}, quantum machine-learning reduced-order modelling for turbulent flows~\cite{li2025quantummachinelearningefficient_3refer}, end-to-end quantum lattice Boltzmann algorithms for nonlinear and non-trivial incompressible flows~\cite{jennings2025endtoendquantumalgorithmnonlinear_2refer,Jennings_2026_6refer}, and HHL-based hybrid quantum--classical solvers for incompressible Navier--Stokes equations~\cite{inger2026hhlbasedquantumclassicalsolverincompressible_4refer}. These works demonstrate the growing relevance of quantum algorithms to CFD, but they also highlight important limitations: learning-based or reduced-order approaches still rely on training data or problem-specific latent spaces, lattice-Boltzmann-based quantum solvers are not directly aligned with conventional stencil-based matrix-free PDE workflows, and HHL-type methods remain centred on global linear-system inversion with state-preparation, conditioning, and readout bottlenecks. Moreover, recent lower-bound results suggest that quantum algorithms should not be expected to provide universal speedups for fluid simulation in full generality~\cite{ameri2026quantumlowerboundssimulating_5refer}. Consequently, a structural gap remains: Neural Physics offers a natural local operator representation for PDEs, but is tied to classical AI hardware; quantum PDE and quantum linear-algebra methods offer compact state and operator manipulation, but do not directly inherit the stencil-centric formulation of classical discretisation.

This gap motivates the present work. We argue that an important missing ingredient in quantum scientific computing is a framework that preserves the analytically derived, local, matrix-free structure of PDE discretisations while enabling quantum state encoding and structured quantum operator implementation. To address this need, we introduce a new framework termed \emph{Quantum Neural Physics} and develop within it a Hybrid Quantum-Classical CNN Multigrid Solver (HQC-CNNMG). The central idea is to map analytically prescribed stencil operators to local quantum convolutional primitives, while retaining classical multilevel scheduling through a W-cycle multigrid architecture. In this way, quantum computation is assigned to structured local operator modules, whereas global iteration control remains classical.

The resulting framework combines three ingredients. From numerical simulation, it inherits the consistency and locality of classical discretisation schemes. From AI, it adopts the operator-centric interpretation of fixed convolutions and the multiscale structural analogy between multigrid and U-Net-like encoder--decoder architectures. From quantum computing, it draws on amplitude encoding for compact state representation and on the Linear Combination of Unitaries (LCU) together with the Quantum Fourier Transform (QFT) for structured operator realisation. In particular, the proposed local quantum convolution operators admit logarithmic-depth circuit implementations with respect to the encoded block size under the idealised circuit model considered in this work.

The contribution of this paper is therefore not the replacement of mature PDE solvers by a fully quantum end-to-end algorithm, nor a claim of universal hardware speedup on current devices. Rather, the main contribution is the establishment of a new algorithmic and architectural pathway for embedding structured quantum local operators into classical multilevel PDE workflows. Within this framework, the key contributions are as follows:
\begin{enumerate}
    \item \textbf{Quantum Neural Physics framework}: We introduce a conceptual bridge between Neural Physics and quantum scientific computing by extending analytically prescribed stencil-based operators from classical convolutional implementations to quantum circuit representations.
    
    \item \textbf{Hybrid multilevel solver architecture}: We develop the HQC-CNNMG framework, in which a classical CPU/GPU manages global W-cycle multigrid scheduling, while selected local operations, including stencil-based convolution and inter-grid transfer, are delegated to quantum subroutines.
    
    \item \textbf{Structured quantum operator construction}: We construct local quantum convolution, restriction, and prolongation operators using amplitude encoding, LCU, and QFT, leading to logarithmic-depth implementations for encoded local blocks under the assumptions of the circuit model employed here.
\end{enumerate}

The proposed framework is assessed on representative PDE benchmarks, including the Poisson equation, transient diffusion equation, convection--diffusion equation, and incompressible Navier--Stokes equations. The aim is to evaluate numerical consistency, multilevel integration, and workflow-level feasibility on noiseless quantum simulators, rather than to claim end-to-end quantum advantage for full classical state reconstruction.

The remainder of this paper is organised as follows. Section~2 presents the methodological foundations of the proposed framework, including the Neural Physics formulation, multigrid architecture, quantum local operator constructions, and the full HQC-CNNMG solver. Section~3 reports numerical experiments on representative benchmark problems. Section~4 compares the proposed approach with representative quantum linear-system paradigms and clarifies its algorithmic position. Section~5 concludes the paper and discusses limitations and future directions.

\section{Methodology}

This section presents the methodology underlying the proposed Hybrid Quantum-Classical CNN Multigrid Solver (HQC-CNNMG). The framework is developed at the intersection of numerical simulation, AI-based operator representation, and quantum computing. From numerical analysis, we adopt PDE discretisation and multigrid as the solver backbone. From AI, we use the Neural Physics viewpoint, in which analytically derived local operators are interpreted as fixed convolutional modules, together with the structural correspondence between multigrid and multiscale encoder--decoder architectures such as U-Net. From quantum computing, we construct compact circuit implementations of these local operators using amplitude encoding, block encoding, the Linear Combination of Unitaries (LCU), and the Quantum Fourier Transform (QFT). The overall development proceeds from local discretisation, to multilevel architecture, to quantum realisation of the resulting operator primitives.

\subsection{Neural Physics and Numerical Discretisation}

The central idea of Neural Physics is that many local numerical discretisations of PDE operators can be recast as fixed convolutional operations. Under this viewpoint, analytically derived stencil coefficients play the role of convolutional kernel weights, so that matrix-free operator application can be implemented through untrained convolutional layers rather than through data-driven parameter learning. This preserves the mathematical structure of the numerical scheme while making local operator evaluation compatible with modern deep-learning software stacks and accelerator hardware.

As a representative example, consider the two-dimensional convection--diffusion equation
\begin{equation}
\frac{\partial T}{\partial t} + \mathbf{v}\cdot\nabla T = D\nabla^2 T + S,
\end{equation}
where $T$ is the scalar concentration field, $\mathbf{v}=(u,v)$ is an incompressible advection velocity satisfying $\nabla\cdot\mathbf{v}=0$, $D$ is the diffusion coefficient, and $S$ is the source term.

The diffusion term is governed by the Laplacian,
\begin{equation}
\nabla^2 T = \frac{\partial^2 T}{\partial x^2} + \frac{\partial^2 T}{\partial y^2},
\end{equation}
which, on a uniform grid with spacing $h$, is approximated by
\begin{equation}
\nabla^2 T_{i,j} \approx \frac{T_{i+1,j}+T_{i-1,j}+T_{i,j+1}+T_{i,j-1}-4T_{i,j}}{h^2}.
\end{equation}
This five-point stencil can be written directly as a fixed $3\times3$ convolution kernel,
\begin{equation}
K_{\text{diff}}=
\frac{1}{h^2}
\left(
\begin{array}{rrr}
0 & 1 & 0 \\
1 & -4 & 1 \\
0 & 1 & 0
\end{array}
\right).
\end{equation}

The convection term is
\begin{equation}
\mathbf{v}\cdot\nabla T = u\frac{\partial T}{\partial x} + v\frac{\partial T}{\partial y}.
\end{equation}
Assuming constant positive velocities $u>0$ and $v>0$, a first-order upwind discretisation yields
\begin{equation}
\mathbf{v}\cdot\nabla T\big|_{i,j}
\approx
u\frac{T_{i,j}-T_{i-1,j}}{h}
+
v\frac{T_{i,j}-T_{i,j-1}}{h}.
\end{equation}
Rearranging the nodal contributions gives
\begin{equation}
\frac{1}{h}\left[(-u)\,T_{i-1,j}+(-v)\,T_{i,j-1}+(u+v)\,T_{i,j}\right],
\end{equation}
which corresponds to the asymmetric convolution kernel
\begin{equation}
K_{\text{conv}}=
\frac{1}{h}
\left(
\begin{array}{ccc}
0 & -u & 0 \\
-v & u+v & 0 \\
0 & 0 & 0
\end{array}
\right).
\end{equation}
For comparison, the second-order central-difference discretisation of the convection term can be written as
\begin{equation}
L_{CD}=
\frac{1}{2h}
\left(
\begin{array}{rrr}
0 & v & 0 \\
u & 0 & -u \\
0 & -v & 0
\end{array}
\right).
\end{equation}

These examples show that both symmetric diffusion operators and asymmetric advection operators admit fixed local convolutional representations. This operator-to-kernel mapping is the basic mechanism underlying the Neural Physics formulation adopted here.

After spatial discretisation, the governing equation can be written in semi-discrete form as
\begin{equation}
\frac{d\mathbf{T}}{dt}= \mathcal{L}(\mathbf{T}),
\end{equation}
where $\mathbf{T}(t)\in\mathbb{R}^N$ collects the nodal values of the scalar field, and
\begin{equation}
\mathcal{L}(\mathbf{T})=-\mathbf{C}\mathbf{T}+D\mathbf{L}\mathbf{T}+\mathbf{S}.
\end{equation}
Here, $\mathbf{C}$ denotes the discrete convection operator, $\mathbf{L}$ the discrete Laplacian operator, and $\mathbf{S}$ the source-term vector.

Applying first-order explicit Euler time integration gives
\begin{equation}
\mathbf{T}^{n+1}
=
\mathbf{T}^{n}
+
\Delta t\left(
-\mathbf{C}\mathbf{T}^{n}
+
D\mathbf{L}\mathbf{T}^{n}
+
\mathbf{S}^{n}
\right),
\end{equation}
or compactly,
\begin{equation}
\mathbf{T}^{n+1}= \mathbf{T}^{n}+ \Delta t\left(\mathbf{A}\mathbf{T}^{n}-\mathbf{b}\right),
\end{equation}
with
\begin{equation}
\mathbf{A}=-\mathbf{C}+D\mathbf{L}, \qquad \mathbf{b}=-\mathbf{S}.
\end{equation}
This compact form collects the local stencil contributions into a single global operator that can be evaluated in matrix-free convolutional form.

At the nodal level, and again assuming $u>0$ and $v>0$, the same update reads
\begin{align}
T_{i,j}^{n+1}
&=
T_{i,j}^{n}
+
\Delta t
\Bigg[
-
\left(
u\frac{T_{i,j}^{n}-T_{i-1,j}^{n}}{h}
+
v\frac{T_{i,j}^{n}-T_{i,j-1}^{n}}{h}
\right)
\nonumber\\
&\qquad
+
D\left(
\frac{T_{i+1,j}^{n}-2T_{i,j}^{n}+T_{i-1,j}^{n}}{h^2}
+
\frac{T_{i,j+1}^{n}-2T_{i,j}^{n}+T_{i,j-1}^{n}}{h^2}
\right)
+
S_{i,j}^{n}
\Bigg].
\end{align}
For spatially varying velocity fields, the same construction applies locally, with coefficients evaluated pointwise. As in standard explicit advection--diffusion discretisations, a sufficient CFL-type stability condition is
\begin{equation}
\Delta t \leqslant \min\left(
\frac{h}{2|u|_{\max}},
\frac{h}{2|v|_{\max}},
\frac{h^2}{4D}
\right).
\end{equation}

From an operator viewpoint, the discrete formulation admits a matrix-free convolutional representation. Instead of assembling a large sparse matrix, the action of the discrete operator is evaluated through compact local kernels:
\begin{equation}
(\mathbf{A}\mathbf{T})_{i,j} = \left(K_{\mathbf A} * T\right)_{i,j},
\end{equation}
where
\begin{equation}
K_{\mathbf A} = -K_{\text{conv}} + D K_{\text{diff}}.
\end{equation}
This matrix-free operator form is the central computational abstraction underlying the Neural Physics formulation.

This representation is not restricted to low-order finite-difference stencils. The same operator-to-kernel viewpoint extends to higher-order local discretisations, including convolutional finite element formulations. For example,
\begin{equation}
K_{\text{Diff\_ConvFEM}}
= \frac{1}{3h^2}
\left(
\begin{array}{rrr}
-1 & -1 & -1 \\
-1 & 8 & -1 \\
-1 & -1 & -1
\end{array}
\right),
\end{equation}
and
\begin{equation}
K_{\text{Conv\_ConvFEM}}
=
\frac{1}{12h}
\left(
\begin{array}{ccc}
u+v & 4v & -u+v \\
4u & 0 & -4u \\
u-v & -4v & -u-v
\end{array}
\right).
\end{equation}
Thus, Neural Physics provides a unified local-operator representation across standard finite-difference and higher-order discretisation schemes. This generality ensures that the subsequent multilevel and quantum constructions inherit a broader operator-centric interface from numerical discretisation.

\subsection{Multigrid Method}

Once the convection and diffusion operators have been discretised, their local stencil-based actions can be evaluated efficiently in matrix-free form. For steady-state problems and implicit time-stepping schemes, however, the dominant task is the solution of the resulting global linear system
\begin{equation}
\mathbf{A}\mathbf{x}=\mathbf{b}.
\end{equation}
This motivates a scalable multilevel solver architecture. The multigrid method is a natural choice.

Multigrid accelerates convergence by exploiting a hierarchy of discretisation levels. High-frequency error components are damped on fine grids by local relaxation, whereas low-frequency errors become more oscillatory on coarser grids and can therefore be corrected more efficiently. Its core operations are smoothing, restriction, prolongation, and coarse-grid correction.

In this work, we adopt a W-cycle strategy, which applies coarse-grid correction more aggressively than the standard V-cycle. Let $\mathcal{G}_h, \mathcal{G}_{2h}, \dots, \mathcal{G}_{Lh}$ denote nested grids from fine to coarse. On each level, the approximate solution is updated through smoothing, residual restriction, coarse-grid correction, and prolongation. The restriction operator $I_h^{2h}$ transfers residuals from $\mathcal{G}_h$ to $\mathcal{G}_{2h}$, while the prolongation operator $I_{2h}^{h}$ maps coarse-grid corrections back to the fine level. We denote by $\mathcal{S}^{\nu}$ the application of $\nu$ smoothing steps. Within the W-cycle, intermediate levels recursively invoke coarse-grid correction twice before returning to finer levels, strengthening low-frequency error reduction.

\begin{figure*}[t]
    \centering
    \includegraphics[width=0.9\textwidth]{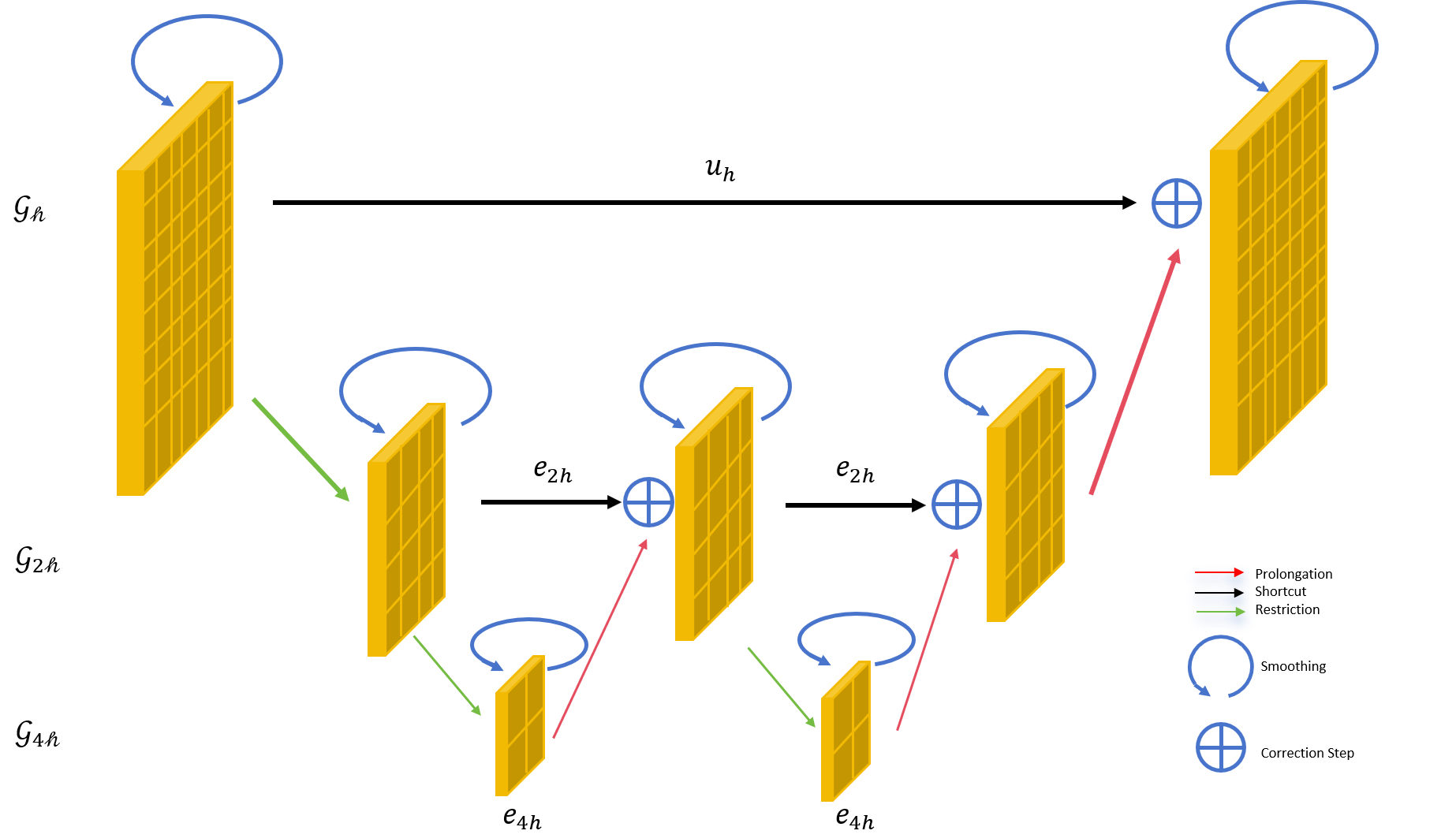}
    \caption{Schematic diagram of the multigrid W-cycle iteration. The figure illustrates the recursive process from the fine grid $\mathcal{G}_h$ to the coarsest grid $\mathcal{G}_{4h}$. The operator actions correspond to the notation introduced in the text and together form a multiscale encoder--decoder-like computational structure reminiscent of U-Net.}
    \label{fig:multi-grid-NN}
\end{figure*}

From the viewpoint of computational structure, multigrid exhibits a natural correspondence with multiscale encoder--decoder architectures such as U-Net~\cite{ronneberger2015unet}. In the present matrix-free setting, local stencil-based operator applications underlying residual evaluation and smoothing can be interpreted as fixed convolutional modules. This does not render multigrid a trainable neural network; rather, it reveals a shared multiscale computational graph in which the operators remain analytically prescribed by numerical discretisation.

At each grid level, local operators act through compact convolutional kernels with fixed weights. Fine-grid and coarse-grid operator applications are analogous to convolutional blocks, while restriction resembles downsampling and prolongation resembles upsampling. Under this view, the W-cycle forms a deeply nested multiscale architecture reminiscent of Nested U-Net~\cite{zhou2018unet++}, with repeated fine-to-coarse and coarse-to-fine information exchange.

This structural correspondence is important for HQC-CNNMG. Multigrid provides the global error-reduction mechanism, while the Neural Physics view provides an interpretable encoder--decoder-like computational backbone. The present framework does not quantise the entire multigrid recursion; instead, it preserves classical global scheduling while assigning quantum computation to selected structured local primitives.

\subsection{Quantum Circuit Representation}

Having established the multigrid hierarchy as the global solver architecture, we now turn to the quantum realisation of its key local building blocks. In the proposed hybrid framework, global multilevel scheduling remains classical, while selected local operators are implemented in quantum circuit form. We consider three classes of primitives: local operator application for stencil-based $\mathbf{A}\mathbf{x}$ evaluation, restriction for coarse-grid transfer, and prolongation for fine-grid correction.

\subsubsection{Amplitude Encoding}

Amplitude encoding embeds classical vector da\-ta into quantum states. Given a normalised vector $\mathbf{x}\in\mathbb{C}^N$ with $N=2^n$, the corresponding amplitude-encoded state is
\begin{equation}
|\mathbf{x}\rangle = \sum_{i=0}^{N-1} x_i |i\rangle,
\end{equation}
where $|i\rangle$ denotes the computational basis state. This encoding requires $n=\log_2 N$ qubits to represent an $N$-dimensional vector. In the present work, it is used to encode local data blocks on which the quantum convolution, restriction, and prolongation operators act.

\subsubsection{Block Encoding}

To implement non-unitary linear operators within quantum circuits, we employ the standard block-encoding framework. For a matrix $A\in\mathbb{C}^{N\times N}$ satisfying $\|A\|\leqslant 1$, a block encoding of $A$ is a unitary operator $U_A$ acting on an enlarged Hilbert space such that
\begin{equation}
(\langle 0^a| \otimes I_N)\, U_A \, (|0^a\rangle \otimes I_N) = A,
\end{equation}
where $a$ denotes the number of ancilla qubits. This provides a unitary representation of a generally non-unitary operator.

In this work, the block encoding is constructed using LCU. For a fixed local convolution stencil, the induced discrete operator $A$ can be decomposed as a weighted sum of translation operators,
\begin{equation}
A = \sum_k c_k\, T_{(d_r^k,d_c^k)},
\end{equation}
where $c_k$ are the stencil coefficients and $T_{(d_r,d_c)}$ denotes the unitary translation associated with shifts by $d_r$ and $d_c$. For an odd-sized kernel of width $s=2m+1$, this decomposition contains $s^2$ translation terms. The LCU construction implements $A/\lambda$, where
\begin{equation}
\lambda = \sum_k |c_k|
\end{equation}
is the normalisation factor.

Translation operators admit an efficient implementation in the Fourier basis. For a one-dimensional translation operator $T_d$,
\begin{equation}
\mathrm{QFT}\cdot T_d \cdot \mathrm{QFT}^\dagger
=
\mathrm{diag}\left(e^{i2\pi k d / N}\right)_{k=0}^{N-1},
\end{equation}
so that translation becomes a phase operation in the Fourier domain. For two-dimen\-sional data, this diagonalisation is applied separately to row and column sub-registers. As a result, the SELECT stage of LCU can be realised through controlled phase rotations, avoiding explicit multi-controlled shift synthesis.

The corresponding block-encoding procedure consists of three stages:
\begin{enumerate}
    \item \textbf{PREPARE}: prepare the ancilla state
    \begin{equation}
    \sum_k \sqrt{|c_k|/\lambda}\, |k\rangle;
    \end{equation}
    \item \textbf{SELECT}: apply the translation-dependent phase operations in the QFT domain;
    \item \textbf{UNPREPARE}: reverse the PREPARE step to complete the block encoding.
\end{enumerate}

Under an idealised circuit model that allows parallel execution of gates acting on disjoint qubits, the translation-based implementation yields polylogarithmic logical depth in the encoded local index registers, with additional cost from coefficient-state preparation. This circuit-level scaling should be distinguished from end-to-end runtime for producing a full classical output vector.

For sparse local operators such as stencil-based convolutions, classical matrix-free evaluation of $\mathbf{A}\mathbf{x}$ already scales linearly with the number of unknowns. Accordingly, the present construction is not intended to claim a universal end-to-end exponential speedup for complete state reconstruction. Rather, it provides a logarithmic-depth local primitive that may be useful for observable estimation or as a subroutine within larger hybrid quantum algorithms.

\subsection{Implementation of Quantum Convolution Circuits}

We consider a fixed odd-sized convolution kernel of width $s=2m+1$ ($m\geqslant 1$), and construct a quantum convolution engine that applies the corresponding local operator to an encoded input block of size $B\times B$, where $B\geqslant s$. For a valid convolution, a single circuit call maps
\begin{equation}
B\times B \longrightarrow (B-s+1)\times(B-s+1).
\end{equation}
The circuit instantiates the amplitude-encoding and LCU--QFT principles in a blockwise form compatible with local stencil evaluation.

The smallest illustrative case is $s=3$ and $B=3$, which gives a $3\times3 \to 1\times1$ valid-convolution unit. The local input contains $9$ grid values and must be embedded into a $2^4=16$ dimensional Hilbert space, leaving unused basis states. Moreover, one circuit call produces only one output value.

\begin{figure*}[t]
    \centering
    \includegraphics[width=\textwidth]{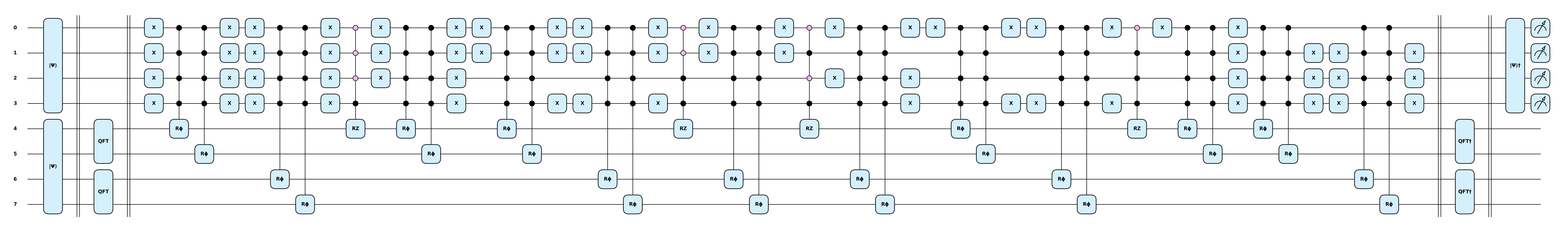}
    \caption{Representative packed-block quantum convolution circuit for the case $s=3$ and $B=4$, corresponding to the mapping $4\times4 \to 2\times2$. The circuit uses 8 qubits in total (4 ancilla qubits and 4 data qubits), with a gate count of 99 and a circuit depth of 53. This example provides a concrete instantiation of the LCU--QFT construction and is consistent with the logarithmic-depth and polylogarithmic gate-complexity scaling of the underlying local operator design.}
    \label{fig:qcnn_circuit}
\end{figure*}

To improve qubit utilisation, we enlarge the encoded input block while keeping the physical stencil fixed. For $s=3$ and $B=4$, one obtains a $4\times4 \to 2\times2$ valid-convolution circuit. The input block contains $16$ grid values, exactly matching the Hilbert space of 4 data qubits, and one circuit call returns a $2\times2$ output patch. This packed-block design improves register utilisation and is more suitable for blockwise sliding-window deployment. Figure~\ref{fig:qcnn_circuit} illustrates this representative construction.

For a kernel of width $s$ acting on a $B\times B$ input block, the induced local convolution operator contains $s^2$ translation terms,
\begin{equation}
A = \sum_{k=1}^{s^2} c_k\, T_{(d_r^k,d_c^k)}.
\end{equation}
The ancilla register for LCU selection scales as $\lceil \log_2(s^2)\rceil$, while the data register contains two sub-registers of size $\lceil \log_2 B\rceil$ for row and column indices. Hence, the total number of qubits per local circuit call scales as
\begin{equation}
O(\log s + \log B).
\end{equation}
Under the idealised parallel circuit model, the logical circuit depth per call is dominated by QFT operations and LCU preparation/unpreparation, yielding
\begin{equation}
O(\log B),
\end{equation}
up to coefficient-state preparation cost.

To apply the local quantum convolution engine to a full input of size $H\times W$, we use a blockwise sliding-window deployment strategy. The domain is partitioned into overlapping or tiled $B\times B$ sub-blocks, and the quantum engine is applied to each complete block to produce a valid output patch. In a tiled valid-output deployment, a natural stride is $B-s+1$, so neighbouring calls generate non-overlapping output patches. Boundary regions that do not admit a complete $B\times B$ block fall back to classical convolution evaluation.

Under this deployment model, the number of local quantum calls scales as
\begin{equation}
O\!\left(\frac{HW}{(B-s+1)^2}\right).
\end{equation}
If these calls are executed sequentially, the corresponding total serial quantum depth scales as
\begin{equation}
O\!\left(\frac{HW}{(B-s+1)^2}\log B\right),
\end{equation}
up to state-preparation costs. This call count characterises the deployment of the local convolution module and does not by itself represent the full runtime of the hybrid multigrid solver.

The blockwise design preserves the functional structure of classical valid convolution while confining quantum computation to local, reusable operator modules. This quantum convolution engine provides the core local primitive used in the construction of quantum restriction, prolongation, and the full HQC-CNNMG solver.

\subsection{Quantum Restriction and Prolongation Circuits}

Having constructed the blockwise quantum convolution engine, we next introduce lightweight quantum transfer operators for multigrid restriction and prolongation. These operators realise local cross-scale actions rather than global transforms.

The \textbf{Quantum Restriction Operator} acts on a local $2\times2$ block of fine-grid data and produces a coarse-grid aggregate. The four local values are amplitude-encoded into a 2-qubit state, after which Hadamard gates are applied to both qubits. The amplitude associated with $|00\rangle$ becomes proportional to the sum of the four encoded components, so the coarse-block aggregate can be extracted through Hadamard interference with appropriate normalisation.

The \textbf{Quantum Prolongation Operator} performs the complementary coarse-to-fine transfer. Starting from a coarse-grid scalar encoded as a local quantum amplitude, two Hadamard gates generate
\begin{equation}
|+\!+\rangle = \frac{1}{2}\left(|00\rangle + |01\rangle + |10\rangle + |11\rangle\right),
\end{equation}
thereby distributing the encoded amplitude uniformly over the four fine-grid basis states. This corresponds to constant local interpolation over a $2\times2$ block.

At the classical operator level, these transfer primitives correspond to averaging-type restriction and piecewise-constant prolongation. Together with quantum convolution, they form a complete set of local operator modules for the proposed hybrid multigrid framework. The quantum processor executes structured low-dimensional primitives, while the classical CPU/GPU remains responsible for global grid scheduling, residual updates, and multilevel iteration control.

\subsection{Hybrid Quantum-Classical CNN Multigrid Solver (HQC-CNN\-MG)}

Building on the multigrid/U-Net structural correspondence and the local quantum operator modules developed above, we now assemble the full HQC-CNNMG solver. The goal is to provide a practical hybrid pathway in which global multilevel control remains classical, while selected local operator applications are delegated to quantum subroutines. The main methodological contribution is the modular substitution of selected local multigrid primitives by structured quantum circuit realisations, rather than the replacement of multigrid scheduling itself.

From an architectural viewpoint, the resulting framework may be interpreted as a physics-encoded quantum-classical analogue of a U-Net in a structural, non-trainable sense. The multigrid hierarchy plays the role of an encoder--decoder-style backbone: restriction acts as downsampling, prolongation acts as upsampling, and local stencil application corresponds to convolution-like transformation with analytically prescribed kernels. The classical CPU/GPU orchestrates global recursion, residual management, and multilevel coupling, while the QPU executes local structured mappings.

Specifically, the HQC-CNNMG framework consists of the following mechanisms:
\begin{itemize}
    \item \textbf{Quantum local operator application}: On each grid level, selected stencil-based local operator evaluations are carried out through the blockwise quantum convolution engine. For a kernel of width $s$ acting on a $B\times B$ local block, each circuit call realises a valid mapping from $B\times B$ to $(B-s+1)\times(B-s+1)$.

    \item \textbf{Quantum inter-grid transfer}: Restriction and prolongation between neighbouring grid levels are implemented through lightweight 2-qubit quantum transfer operators, providing local coarse-grid aggregation and fine-grid redistribution.

    \item \textbf{Hybrid blockwise deployment and fallback}: Quantum modules are invoked only on complete local blocks compatible with the prescribed register size. Near boundaries, or whenever the local block configuration is unsuitable, the algorithm falls back to the corresponding classical operator without altering the global multigrid logic.

    \item \textbf{Classical W-cycle scheduling}: The overall solver adopts the W-cycle mul\-tigrid strategy. Recursion, residual management, coarse-grid correction scheduling, and stopping control remain classical, while local operator applications are replaced where appropriate by quantum subroutines.
\end{itemize}

The computational structure of HQC-CNNMG is hierarchical. At the level of a single quantum convolution call, the module requires $O(\log s+\log B)$ qubits and has logical depth $O(\log B)$, up to the LCU state-preparation cost. The restriction and prolongation modules act on fixed $2\times2$ blocks and remain constant-depth local primitives.

At a multigrid level with $N_\ell$ unknowns, the number of blockwise convolution calls in the tiled valid-output setting scales as
\begin{equation}
O\!\left(\frac{N_\ell}{(B-s+1)^2}\right),
\end{equation}
with corresponding serial quantum depth proportional to this call count times the depth of each local convolution call. The transfer operators contribute only lower-order local overhead.

From the global solver perspective, classical multigrid scheduling retains the linear-work structure of geometric multigrid, while selected local operator applications are replaced by logarithmic-depth quantum subroutines. Therefore, the proposed framework should be understood as a hybrid multilevel cost model rather than as a claim of universal end-to-end asymptotic improvement for the full PDE solve. Its value lies in providing a structured pathway for embedding quantum local operators into matrix-free PDE workflows while preserving the stability and multiresolution correction mechanisms of classical multigrid.

\section{Experiments}

This section evaluates the proposed Hybrid Sliding Window Scheme for local forward operator application and the Hybrid Quantum-Classical CNN Multigrid solver (HQC-CNNMG) on steady-state and time-dependent PDE problems. The experiments are organised in increasing order of complexity: operator-level verification, elliptic and transient PDE benchmarks, and finally a nonlinear incompressible-flow demonstration. This progression assesses numerical consistency, multilevel solver accuracy, temporal robustness, and workflow-level feasibility.

\subsection{Experimental Setup}

\paragraph{Environment and implementation.}
All experiments were conducted on a workstation equipped with 2 $\times$ Intel(R) Xeon(R) CPU E5-2699 v3 @ 2.30GHz processors and 1 $\times$ NVIDIA GeForce RTX 2080 Ti GPU with 11 GB VRAM. The quantum components, including local quantum convolution, restriction, and prolongation modules, were implemented in \texttt{PennyLane}~\cite{bergholm2018pennylane} and embedded into the hybrid multigrid workflow.

\paragraph{Simulator and reference protocol.}
All quantum results were obtained using the noiseless \texttt{default.qubit} state-vector simulator in \texttt{PennyLane}. The experiments therefore evaluate algorithmic correctness and workflow feasibility under ideal circuit execution, rather than noisy-hardware performance. Unless otherwise stated, reference solutions were generated classically using analytical solutions or high-accuracy sparse direct solvers.

\paragraph{Metrics and stopping criteria.}
The evaluation uses relative error, maximum absolute error, and $L^2$ error, together with problem-specific quantities such as peak amplitude, centroid displacement, total mass discrepancy, and pressure residual. For elliptic and implicit subproblems, convergence is assessed by comparison with reference solutions after the prescribed number of W-cycle applications. The emphasis is on numerical verification rather than wall-clock benchmarking.

\subsection{Basic Verification}

\subsubsection{Linear System: Operator Correctness}

We first verify the matrix-free hybrid formulation by solving a linear system derived from the standard five-point discretisation of the two-dimensional Poisson operator,
\begin{equation}
\mathbf{A}\mathbf{x}=\mathbf{b},
\end{equation}
where $\mathbf{A}\in\mathbb{R}^{N\times N}$ is the discrete Laplacian matrix and $N=H\times W$. This test examines whether the local quantum convolution module reproduces the corresponding classical stencil action within the HQC-CNNMG iteration.

For reference generation, the sparse matrix $\mathbf{A}$ was explicitly assembled using the five-point stencil with coefficients $[4,-1,-1,-1,-1]$ at interior nodes:
\begin{equation}
\begin{split}
\mathbf{A} ={} & \mathrm{diag}(4,0)+\mathrm{diag}(-1,-1)+\mathrm{diag}(-1,1) \\
& +\mathrm{diag}(-1,-W)+\mathrm{diag}(-1,W),
\end{split}
\end{equation}
with horizontal couplings at row boundaries removed to enforce homogeneous Dirichlet conditions. The right-hand side is a smooth Gaussian source,
\begin{equation}
b_{i,j}=10\exp\left(-\frac{(i-H/2)^2+(j-W/2)^2}{15}\right).
\end{equation}
Two grid resolutions are considered,
\begin{equation}
(H,W)=(16,32), \qquad (24,48),
\end{equation}
corresponding to $N=512$ and $N=1152$. Reference solutions are computed using the sparse direct solver \texttt{spsolve}.

In HQC-CNNMG, the global matrix $\mathbf{A}$ is not explicitly applied. Instead, its action is evaluated in matrix-free form as
\begin{equation}
(\mathbf{A}T)_{i,j}=\left(K_{\mathbf A}*T\right)_{i,j},
\end{equation}
where $K_{\mathbf A}$ is the discrete Laplacian kernel. This local operation is carried out through the proposed $4\times4 \to 2\times2$ hybrid sliding-window quantum convolution module.

\begin{figure}[htbp]
    \centering
    \includegraphics[width=\linewidth]{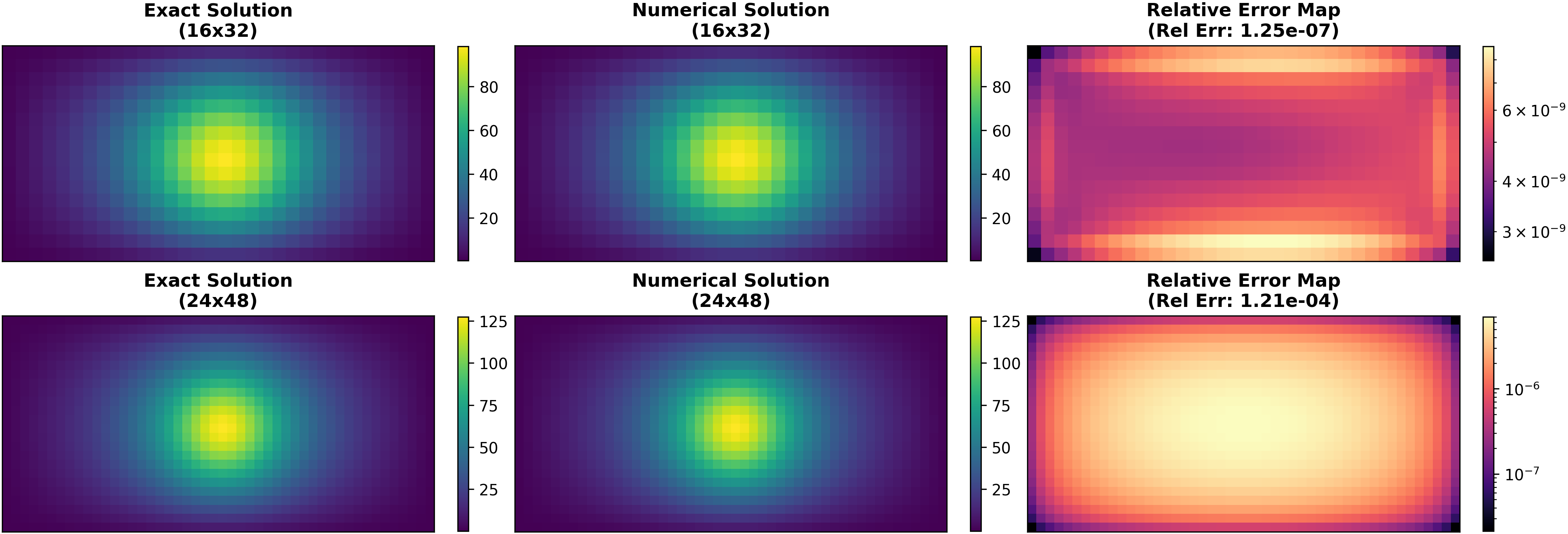}
    \caption{Operator-level verification for the linear system $\mathbf{A}\mathbf{x}=\mathbf{b}$. The first row corresponds to the $16\times32$ grid and the second row to the $24\times48$ grid. From left to right, the panels show the classical reference solution obtained by \texttt{spsolve}, the HQC-CNNMG solution, and the corresponding spatial distribution of the relative error.}
    \label{fig:Ax}
\end{figure}

Figure~\ref{fig:Ax} compares the hybrid and classical solutions. For the $16\times32$ and $24\times48$ grids, the maximum relative errors are $1.25\times10^{-7}$ and $1.21\times10^{-4}$, respectively. These results confirm that the matrix-free hybrid implementation reproduces the intended stencil-based forward operator with high accuracy.

\subsubsection{Poisson: Elliptic Solver Accuracy}

We next assess the full HQC-CNNMG solver on the two-dimensional Poisson equation
\begin{equation}
-\nabla^2 \phi(x,y)=\rho(x,y), \qquad (x,y)\in\Omega=[0,L_x]\times[0,L_y],
\end{equation}
discretised by the standard second-order central difference scheme on a $24\times40$ grid, corresponding to $N=960$ unknowns. A zero initial guess is used,
\begin{equation}
\phi^{(0)}(x,y)=0,
\end{equation}
with homogeneous Dirichlet boundary conditions on $\partial\Omega$.

The source term is a dipole-like charge distribution,
\begin{equation}
\begin{split}
\rho(x,y)={}&20\exp\left(-\frac{(x-x_+)^2+(y-y_+)^2}{15}\right)\\
&-20\exp\left(-\frac{(x-x_-)^2+(y-y_-)^2}{15}\right),
\end{split}
\end{equation}
where $(x_+,y_+)=(L_x/3,L_y/3)$ and $(x_-,y_-)=(2L_x/3,L_y/3)$.

\begin{figure}[htbp]
    \centering
    \includegraphics[width=\linewidth]{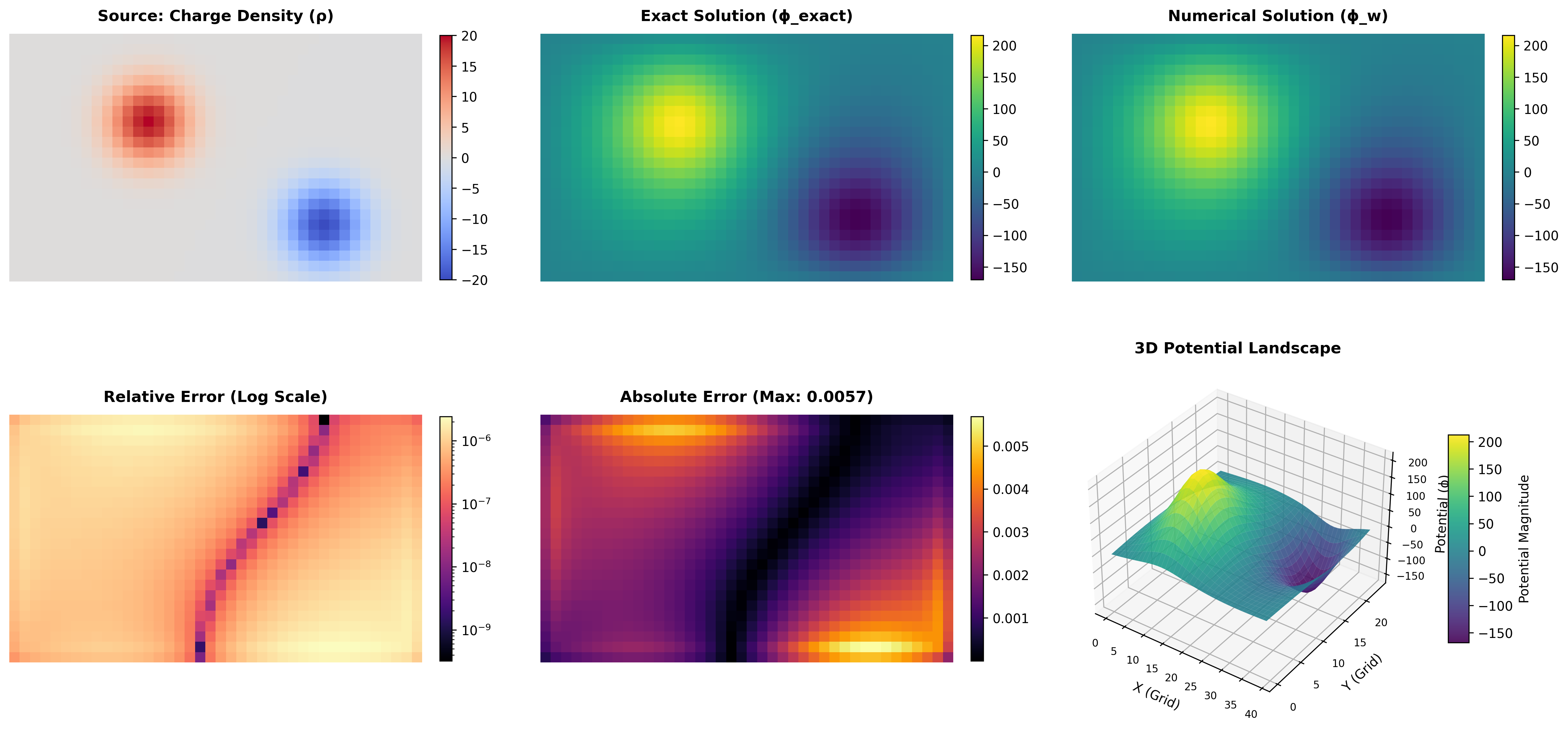}
    \caption{Elliptic benchmark for the Poisson equation on a $24\times40$ grid. The panels show the source term, the classical reference solution, the HQC-CNNMG solution, and the associated relative- and absolute-error fields, together with the reconstructed potential distribution.}
    \label{fig:wcycle_poisson}
\end{figure}

Figure~\ref{fig:wcycle_poisson} shows the source term, reference solution, HQC-CNNMG solution, and error fields. After 6 W-cycle iterations, the global relative error is of order $10^{-6}$ and the maximum absolute error is $0.0057$, indicating accurate elliptic reconstruction across both local gradients and the global potential field.

\subsection{Time-Dependent Validation}

\subsubsection{Diffusion: Repeated Implicit Solve Stability}

We then evaluate repeated implicit time stepping for the transient diffusion equation
\begin{equation}
\frac{\partial \phi}{\partial t}=\alpha\nabla^2\phi+\rho(x,y), \qquad (x,y)\in\Omega=[0,L_x]\times[0,L_y], \quad t>0,
\end{equation}
with diffusion coefficient $\alpha=1$. Time integration uses implicit Euler with $\Delta t=0.5$ over 20 time steps, and the spatial discretisation uses the standard second-order five-point Laplacian on a $16\times24$ grid, corresponding to $N=384$ unknowns.

The initial condition is
\begin{equation}
\phi(x,y,0)=0,
\end{equation}
with homogeneous Dirichlet boundary conditions. The source term is
\begin{equation}
\rho(x,y)=10\exp\left(-\frac{(x-L_x/2)^2+(y-L_y/2)^2}{15}\right),
\end{equation}
which acts as a localised heat source.

\begin{figure}[htbp]
    \centering
    \includegraphics[width=\linewidth]{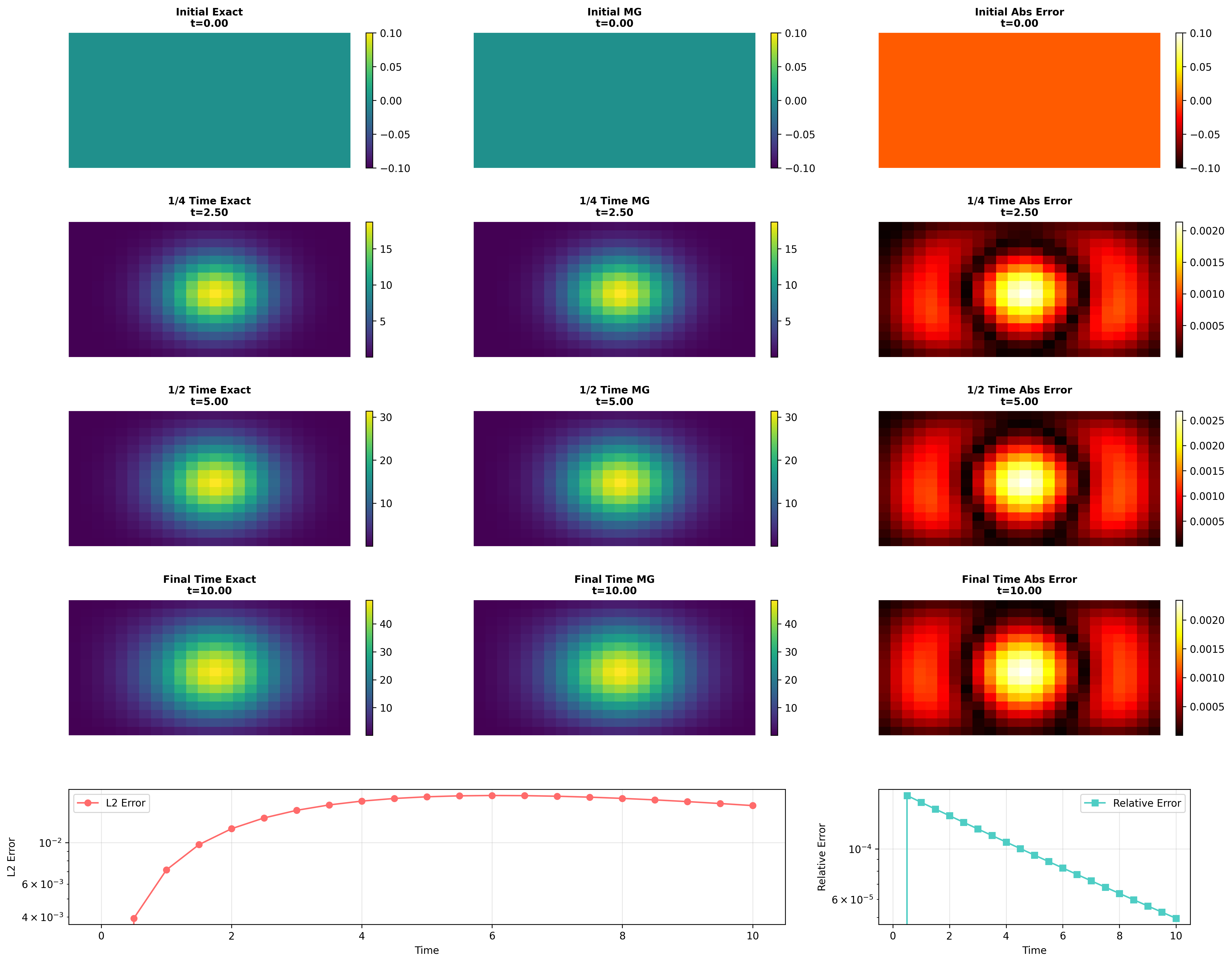}
    \caption{Transient diffusion benchmark on a $16\times24$ grid. The upper panels compare the classical reference solutions, HQC-CNNMG solutions, and absolute-error fields at representative time steps. The bottom panels report the temporal evolution of the $L^2$ error and relative error under repeated implicit solves.}
    \label{fig:diffusion_comparison}
\end{figure}

Figure~\ref{fig:diffusion_comparison} compares the reference and HQC-CNNMG solutions over time. Across all 20 steps, the solver remains stable under repeated W-cycle applications. The $L^2$ error remains at the order of $10^{-2}$, while the relative error decreases to the order of $10^{-5}$. The largest errors occur near the source region, where the gradients are strongest.

\subsubsection{Convection-Diffusion: Physical Transport Fidelity}

To test trans\-port-dominated behaviour, we consider the two-dimensional transient convection--diffusion equation
\begin{equation}
\frac{\partial \phi}{\partial t}
=
\alpha\nabla^2\phi
-
\left(
u\frac{\partial \phi}{\partial x}
+
v\frac{\partial \phi}{\partial y}
\right),
\qquad (x,y)\in\Omega=[0,L_x]\times[0,L_y],
\end{equation}
with $\alpha=0.01$, $u=0.3\,\mathrm{m/s}$, and $v=0.15\,\mathrm{m/s}$. The diffusion and convection terms use the second-order ConvFEM-based local operators introduced in Section~2. Time integration is performed by implicit Euler with $\Delta t=0.01\,\mathrm{s}$ up to $T=1\,\mathrm{s}$ on a $32\times32$ grid over a $2\,\mathrm{m}\times2\,\mathrm{m}$ domain.

The initial condition is a Gaussian pulse,
\begin{equation}
\phi(x,y,0)=\exp\left(-\frac{(x-x_0)^2+(y-y_0)^2}{2\sigma_0^2}\right),
\end{equation}
with $(x_0,y_0)=(0.5,0.5)$ and $\sigma_0=0.12$. No source term is applied, and homogeneous Dirichlet boundaries provide an accurate finite-domain approximation because the pulse remains well separated from the boundaries.

\begin{figure*}[htbp]
    \centering
    \includegraphics[width=\linewidth]{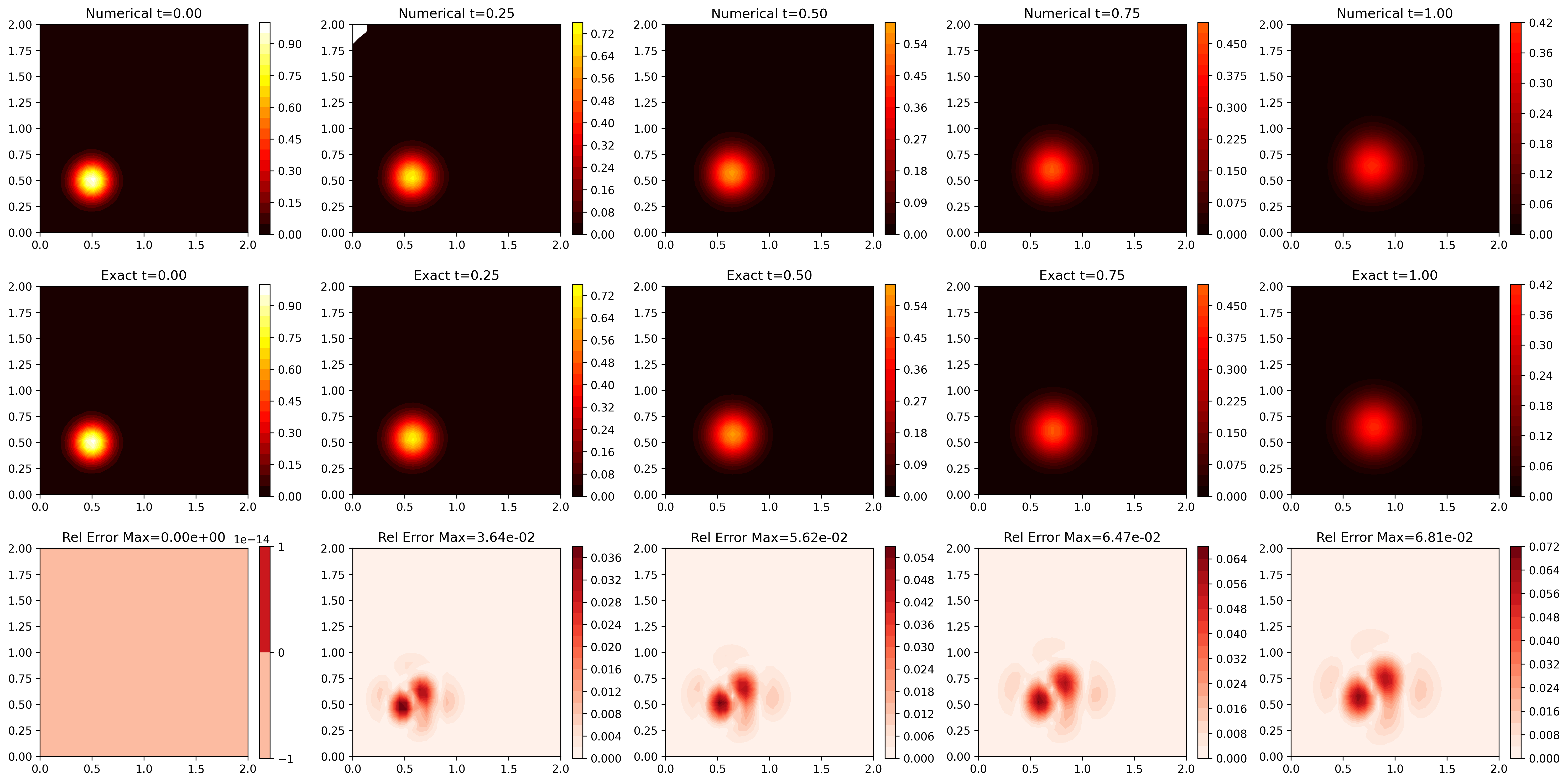}
    \caption{Field-level validation of the convection--diffusion benchmark on a $32\times32$ grid. The panels compare analytical and HQC-CNNMG solutions at representative time steps together with the corresponding relative-error fields.}
    \label{fig:convection_diffusion_evolution}
\end{figure*}

\begin{figure*}[htbp]
    \centering
    \includegraphics[width=\linewidth]{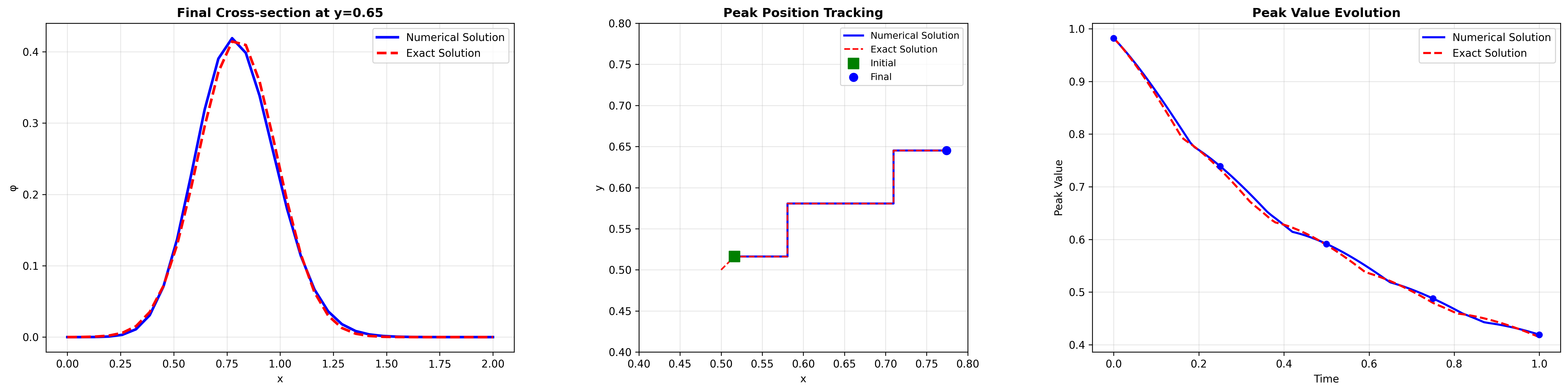}
    \caption{Reduced-quantity validation of the convection--diffusion benchmark on a $32\times32$ grid. The panels compare the final-time solution profiles, the trajectory of the pulse centre, and the temporal attenuation of the peak amplitude, thereby assessing the preservation of key transport observables.}
    \label{fig:convection_diffusion_quantitative}
\end{figure*}

Figures~\ref{fig:convection_diffusion_evolution} and \ref{fig:convection_diffusion_quantitative} compare the numerical and analytical solutions. The relative error remains at the order of $10^{-2}$ throughout the simulation. At final time, the numerical pulse centre is $(x_{\mathrm{num}},y_{\mathrm{num}})=(0.800,0.650)$, agreeing with the theoretical position to three decimal places. The numerical peak value is $0.4143$, compared with the analytical value $0.4144$, and the final mass discrepancy is $0.0673\%$. These results indicate that the solver preserves convective displacement, diffusive spreading, and integral mass behaviour.

\subsection{Complex Flow Demonstration}

\subsubsection{Navier--Stokes: Proof-of-Concept Workflow Integration}

As a final experiment, we consider two-dimensional incompressible flow past a square cylinder to test integration into a nonlinear CFD workflow with advection, diffusion, and pressure--velocity coupling. The governing equations are
\begin{equation}
\begin{aligned}
\frac{\partial \mathbf{u}}{\partial t}+(\mathbf{u}\cdot\nabla)\mathbf{u}
&=
-\nabla p+\nu\nabla^2\mathbf{u}, \\
\nabla\cdot\mathbf{u}&=0,
\end{aligned}
\end{equation}
where $\mathbf{u}=(u,v)$, $p$ is pressure, and $\nu=0.1$. The domain is a $256\times64$ grid with physical size $256\,\mathrm{m}\times64\,\mathrm{m}$. The inflow velocity is $U=1\,\mathrm{m/s}$, the square-cylinder width is $L=12\,\mathrm{m}$, and the Reynolds number is $\mathrm{Re}=UL/\nu=120$. Time integration uses explicit Euler with $\Delta t=0.05\,\mathrm{s}$.

The convection and diffusion terms are represented by the ConvFEM-based local operator formulation introduced in Section~2. Pressure--velocity coupling is handled by the SIMPLE algorithm, and the pressure Poisson subproblem is solved using HQC-CNNMG at each time step. Thus, the proposed hybrid pressure solver is embedded directly into the incompressible-flow update loop.

The initial conditions are
\begin{equation}
\mathbf{u}(x,y,0)=\mathbf{0}, \qquad p(x,y,0)=0.
\end{equation}
The square obstacle is represented by an immersed-boundary damping term $\sigma=10^8$ inside the solid region. Boundary conditions include uniform inflow at the inlet, homogeneous Neumann conditions at the outlet, no-slip top and bottom walls, and a fixed pressure reference at the top-right corner.

The pressure residual decreases from $1.084\times10^{-2}$ at time step 3000 to $1.57\times10^{-3}$ at step 6000 and $2.1\times10^{-4}$ at step 9000, before remaining at the order of $10^{-4}$ or below.

\begin{figure*}[t]
    \centering
    \includegraphics[width=0.95\textwidth]{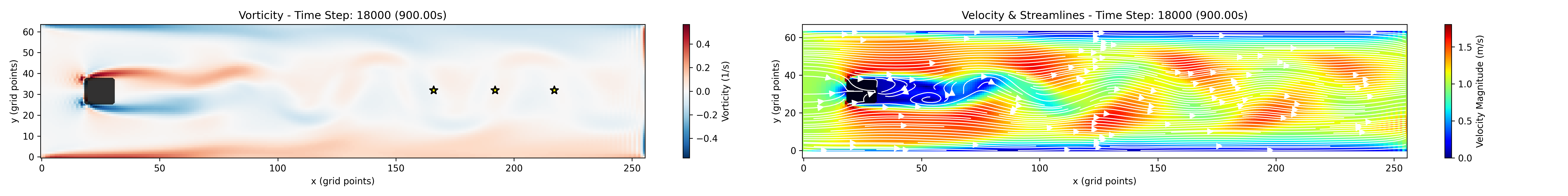}
    \caption{Flow past a square cylinder at $\mathrm{Re}=120$ and $t=900\,\mathrm{s}$. Left: vorticity field showing alternating wake shedding behind the obstacle. Right: velocity magnitude, velocity vectors, and streamlines, illustrating the unsteady recirculating wake structure consistent with a K\'arm\'an vortex street.}
    \label{fig:navier_stokes_block_flow}
\end{figure*}

Figure~\ref{fig:navier_stokes_block_flow} shows the flow field at $t=900\,\mathrm{s}$, where a clear unsteady wake has formed behind the cylinder. The vorticity field exhibits alternating vortex shedding, and the streamline pattern is consistent with a K\'arm\'an vortex street at moderate Reynolds number. This case is intended as a proof-of-concept demonstration rather than a complete CFD benchmark, showing that HQC-CNNMG can be integrated into a nonlinear incompressible-flow workflow while maintaining low pressure residuals and reproducing the expected large-scale wake structure.

\section{Comparison with Representative Quantum Linear Solver Paradigm}

This section positions the proposed Hybrid Quantum-Classical CNN Multigrid framework (HQC-CNNMG) within representative quantum linear solver paradigms. The purpose is not to claim universal superiority, but to clarify how a PDE-oriented hybrid architecture differs from global quantum inversion and optimisation-based approaches when applied to discretized elliptic-type systems.

We compare HQC-CNNMG with four representative paradigms: the Variational Quantum Linear Solver (VQLS)~\cite{bravo2023VQLS}, the HHL algorithm~\cite{harrow2009HHL}, Quantum Singular Value Transformation (QSVT)-based solvers~\cite{QSVT2019Gily}, and Quadratic Unconstrained Binary Optimization (QUBO) formulations with adiabatic evolution~\cite{Lucas2014QUBO}. These methods differ in data-access assumptions, output models, and dominant bottlenecks. The comparison therefore focuses on threshold attainment, relative error, algorithmic overhead, and compatibility with PDE structure, rather than on a single scalar ranking.

\subsection{Comparison Protocol}

The benchmark problem is the linear system
\begin{equation}
\mathbf{A}\mathbf{x}=\mathbf{b},
\end{equation}
with $N=16$, $\mathbf{A}\in\mathbb{R}^{16\times16}$, and $\mathbf{x},\mathbf{b}\in\mathbb{R}^{16\times1}$. This corresponds to a discretized elliptic subproblem on a two-dimensional $4\times4$ grid, where the local operator is instantiated by a standard Laplacian stencil or by a nine-point convolution stencil. Although small, this benchmark is sufficient to expose the different numerical behaviours of the compared paradigms under controlled conditions.

All methods are tested under the same target tolerance,
\begin{equation}
\epsilon < 10^{-2},
\end{equation}
where $\epsilon$ denotes the relative solution error. Simulations are carried out on a PennyLane state-vector simulator, with SciPy and PyTorch used for state preparation, sampling, optimisation, and auxiliary numerical routines. Since simulator wall-clock time is not directly representative of physical quantum-hardware runtime, absolute runtime is not used as the main metric. Instead, we report whether each method reaches the target threshold, its mean relative error, and its iteration or cycle count.

The benchmark should therefore be interpreted as a controlled numerical comparison, rather than as a definitive hardware assessment. Its main role is to examine whether assigning quantum computation to local structured subroutines and embedding them into a classical multigrid hierarchy provides a balanced alternative to fully quantum global inversion.

\subsection{Theoretical Resource Scaling}

For a PDE-discretized system of global size $N$, different quantum linear solver paradigms have different scaling mechanisms. HHL and QSVT are global spectral-transformation methods, whose costs depend on full-system quantities such as $\log N$, the condition number $\kappa$, block-encoding quality, and target precision. VQLS reduces circuit depth through a variational ansatz, but introduces optimisation and measurement overhead. QUBO-based methods convert continuous variables into binary optimisation variables, which increases representation cost and may lead to unfavourable adiabatic scaling.

By contrast, HQC-CNNMG adopts a local multiresolution strategy. Its quantum subroutines act on local receptive fields of size $K\times K$, while global error transport is handled by a classical W-cycle through smoothing, restriction, prolongation, and coarse-grid correction. Thus, the relevant quantum cost is the repeated application of local structured operators, rather than the spectral inversion of the full system.

Table~\ref{tab:theory_complexity} summarises the leading-order resource requirements of the compared paradigms. Here, $K$ denotes the local block size, $\kappa$ the condition number, $\epsilon$ the target relative error, $L$ the VQLS ansatz depth, $n_{clk}$ the number of HHL clock-register qubits, $s$ the matrix sparsity, and $c$ a constant associated with adiabatic spectral-gap scaling. The entries are paradigm-level estimates under their respective standard assumptions and are not directly comparable as end-to-end costs for full classical vector reconstruction.

\begin{table*}[htbp]
\centering
\tiny
\caption{Paradigm-level leading-order resource scaling for representative quantum linear solver approaches applied to an $N$-dimensional discretized system. For HQC-CNNMG, $K$ denotes the side length of the local convolution block or sliding window.}
\label{tab:theory_complexity}

\begin{threeparttable}
\setlength{\tabcolsep}{2.0pt}
\renewcommand{\arraystretch}{1.20}

\begin{tabularx}{\textwidth}{@{}
>{\raggedright\arraybackslash}p{0.17\textwidth}
>{\centering\arraybackslash}p{0.13\textwidth}
>{\centering\arraybackslash}p{0.23\textwidth}
>{\centering\arraybackslash}p{0.15\textwidth}
>{\centering\arraybackslash}p{0.13\textwidth}
>{\centering\arraybackslash}p{0.14\textwidth}
@{}}
\toprule
\textbf{Algorithm Model}
& \textbf{Space Complexity}
& \textbf{Total Gate Count}
& \textbf{Physical Depth}
& \textbf{Preprocessing / Encoding Cost}
& \textbf{Observable / Readout Cost} \\
\midrule

\textbf{VQLS}
& $\mathcal{O}(\log N)$
& $\mathcal{O}(\mathrm{poly}(\kappa)\log(1/\epsilon)\log N)$
& $\mathcal{O}(L)$
& $\mathcal{O}(\log N)$
& $\mathcal{O}(1/\epsilon^2)$ \\

\textbf{HHL}
& $\mathcal{O}(\log N)+1_{\mathrm{anc}}+n_{\mathrm{clk}}$
& $\mathcal{O}(\log N \cdot s^2\kappa^2/\epsilon)$
& $\mathcal{O}(s^2\kappa^2/\epsilon)$
& $\mathcal{O}(\log N)$
& $\mathcal{O}(1)$ \\

\textbf{QSVT}
& $\mathcal{O}(\log N)+\mathcal{O}(\log N)_{\mathrm{anc}}$
& $\mathcal{O}(\kappa\log(1/\epsilon)\,\mathrm{polylog}(N))$
& $\mathcal{O}(\kappa\log(1/\epsilon))$
& $\mathcal{O}(\log N)$
& $\mathcal{O}(1)$ \\

\textbf{QUBO}
& $\mathcal{O}(N\log(1/\epsilon))$
& $\mathcal{O}(e^N)$
& $\mathcal{O}(e^{cN\log(1/\epsilon)})$
& $\mathcal{O}(N^2)$
& $\mathcal{O}(1)^{\dagger}$ \\

\makecell[l]{\textbf{HQC-CNNMG}\\\textbf{(Block Encoding)}}
& $\mathcal{O}(\log K^2)+1_{\mathrm{anc}}$
& $\mathcal{O}\!\left(\frac{N}{K^2} C_{\mathrm{BE}}(K,\epsilon_{\mathrm{loc}})\right)$
& $\mathcal{O}(D_{\mathrm{BE}}(K))$
& $\mathcal{O}(K^2)$
& $\mathcal{O}(K^2/\epsilon_{\mathrm{obs}}^2)$ \\

\makecell[l]{\textbf{HQC-CNNMG}\\\textbf{(LCU-QFT)}}
& $\mathcal{O}(\log K^2)+4_{\mathrm{anc}}$
& $\mathcal{O}\!\left(\frac{N}{K^2} C_{\mathrm{LCU}}(K,\epsilon_{\mathrm{loc}})\right)$
& $\mathcal{O}(\log K)$
& $\mathcal{O}(K^2)$
& $\mathcal{O}(K^2/\epsilon_{\mathrm{obs}}^2)$ \\

\bottomrule
\end{tabularx}

\begin{tablenotes}[flushleft]
\tiny
\item[$\dagger$] A single readout of the final spin configuration on QUBO hardware may be viewed as $\mathcal{O}(1)$, but repeated sampling may be required in practice because the minimum spectral gap can shrink rapidly with system size. For HQC-CNNMG, $C_{\mathrm{BE}}$ and $C_{\mathrm{LCU}}$ denote the local circuit cost of one blockwise operator application. The end-to-end hybrid solver cost also depends on deployment count, classical multigrid scheduling, and stopping criteria.
\end{tablenotes}

\end{threeparttable}
\end{table*}

The main distinction is that VQLS, HHL, QSVT, and QUBO are organised around global variational optimisation, spectral inversion, or discrete optimisation, whereas HQC-CNNMG restricts quantum computation to local PDE operators. Wit\-h\-in HQC-CNNMG, the LCU-QFT realization is particularly attractive because it diagonalises translation operators in the Fourier basis and reduces the local physical depth to $\mathcal{O}(\log K)$ under the idealised parallel circuit model.

\subsection{Benchmark Results and Interpretation}

To evaluate empirical behaviour, we perform three independent runs for each method under the same simulator environment, target tolerance, and input setting. The averaged results are reported in Table~\ref{tab:empirical_benchmark}.

\begin{table*}[htbp]
\centering
\tiny
\caption{Benchmark convergence statistics for the $N=16$ linear-system test, averaged over three independent runs. A maximum iteration cap is imposed on VQLS, while QUBO uses finite-bitwidth approximations for continuous variables together with heuristic sampling.}
\label{tab:empirical_benchmark}

\setlength{\tabcolsep}{3pt}
\renewcommand{\arraystretch}{1.20}

\begin{tabularx}{0.95\textwidth}{
>{\raggedright\arraybackslash}p{0.32\textwidth}
>{\centering\arraybackslash}p{0.18\textwidth}
>{\centering\arraybackslash}p{0.18\textwidth}
>{\centering\arraybackslash}p{0.17\textwidth}
}
\toprule
\textbf{Solver Model}
& \textbf{Reached Target Threshold?}
& \textbf{Mean Relative $L^2$ Error}
& \textbf{Mean Iterations / Cycles} \\
\midrule

\textbf{VQLS}
& Yes
& 0.009016
& 59.0 \\

\textbf{HHL}
& No
& 0.077856
& 1.0 \\

\textbf{QSVT}
& Yes
& $<10^{-6}$
& 1.0 \\

\textbf{QUBO}
& No
& 0.662328
& 1.0 \\

\makecell[l]{\textbf{HQC-CNNMG}\\\textbf{(Block Encoding)}}
& Yes
& 0.000080
& 6.0 \\

\makecell[l]{\textbf{HQC-CNNMG}\\\textbf{(LCU-QFT)}}
& Yes
& 0.000080
& 6.0 \\

\bottomrule
\end{tabularx}
\end{table*}

Table~\ref{tab:empirical_benchmark} shows three main behaviours. QUBO and HHL do not reach the prescribed error threshold. VQLS reaches the threshold, but requires a relatively large number of classical--quantum feedback iterations. QSVT and both HQC-CNNMG variants reach the target with substantially smaller final error.

The poor QUBO result is mainly due to the finite-bitwidth encoding of continuous elliptic unknowns, which introduces an additional discretisation layer and converts the original linear system into a combinatorial optimisation problem. VQLS avoids this binary representation but shifts the burden to ansatz design, parameter optimisation, and measurement-based objective estimation. HHL is sensitive to finite clock-register resolution and inversion precision in the present simulator setting, whereas QSVT provides a high-accuracy global inversion baseline when block encoding is well controlled.

Against this background, HQC-CNNMG exhibits a different type of strength. Both its block-encoding and LCU-QFT variants converge to a mean relative $L^2$ error of approximately $8\times10^{-5}$ within 6 multigrid cycles. Although this does not surpass QSVT in raw small-scale accuracy, it improves substantially over VQLS and HHL in this benchmark while avoiding the representational mismatch of QUBO. More importantly, it supports the main design principle of the framework: quantum computation is assigned to structured local subroutines, such as convolution, restriction, and prolongation, while global error propagation remains classical.

The two HQC-CNNMG variants achieve the same numerical accuracy and cycle count in this small test. Their difference lies primarily in circuit-level realizability. By implementing spatial translations as diagonal phase rotations in the Fourier domain, the LCU-QFT construction yields a shallower local circuit and is therefore more attractive when circuit depth is a dominant implementation constraint.

Overall, these results suggest that HQC-CNNMG should be viewed not as another fully quantum global linear solver, but as a PDE-oriented hybrid architecture. Its contribution lies in combining local quantum operator primitives with classical multilevel error correction, thereby preserving stencil locality, matrix-free structure, and compatibility with existing scientific-computing workflows.

\section{Conclusion}

This paper introduced a \emph{Quantum Neural Physics} framework and, within it, developed a Hybrid Quantum-Classical CNN Multigrid Solver (HQC-CNNMG) for PDE-oriented scientific computing. The central contribution is a reorganisation of the computational workflow in which local PDE discretisations are recast as fixed convolutional operators, embedded within a multilevel multigrid/U-Net-like architecture, and selectively realised as structured quantum circuit primitives based on LCU and QFT. Under this formulation, quantum computation is assigned to local structured operators, while global multilevel coordination remains classical.

At the circuit level, the LCU-QFT construction yields logarithmic-depth local operator implementations, with depth scaling as $\mathcal{O}(\log K)$ under the idealised parallel circuit model considered here. The numerical results support this hybrid interpretation: across operator-level verification, elliptic and transient benchmarks, and a nonlinear incompressible Navier--Stokes workflow, the method demonstrates numerical consistency, stable multilevel behaviour, and compatibility with representative PDE applications. Comparisons with representative quantum linear solver paradigms further indicate that the principal strength of HQC-CNNMG lies not in fully quantum global inversion, but in its balanced trade-off among local circuit depth, numerical robustness, and compatibility with PDE structure.

All results in this work were obtained on noiseless state-vector simulators, and no end-to-end hardware speedup is claimed. State preparation, data movement, and measurement remain major practical bottlenecks. The present study should therefore be interpreted as an algorithmic and architectural feasibility investigation. Future work should extend the framework toward more hardware-realistic and more deeply quantum-integrated multilevel PDE workflows, including noisy and hardware-constrained implementations of the proposed local operators, improved strategies for state preparation and observable extraction, and the development of more fully quantum multilevel components such as quantum-native inter-grid transfer and coarse-level correction. Rather than simply replacing the classical multigrid backbone, such developments would amount to a progressive quantumisation of PDE solver workflows guided by operator locality, multiresolution structure, and scientific-computing requirements. It is also of interest to extend the framework to higher-order discretisations, three-dimensional problems, and coupled multiphysics systems, thereby further clarifying the role of Quantum Neural Physics as an interface between scientific machine learning, numerical simulation, and quantum computing.

\section*{Funding sources}
The authors are grateful to Sinopec as sponsors of the Resource Geophysics Academy, Imperial College London, for supporting this research. The authors would like to acknowledge the following EPSRC grants: AI-STARS, ``Debris transportation based on AI modelling for nuclear site decommissioning - AI-STARS: AI for Investigating Sludge Transport And RadWaste Removal'' (UKRI3252); the PREMIERE programme grant, ``AI to enhance manufacturing, energy and healthcare'' (EP/T000414/1); AI4Urban-Health, ``AI Solutions to Urban Health Using a Place-Based Approach'' (APP55547 / UKRI1241); GP4Streets, ``Pioneering Climate Adaptation in Urban Environments'' (UKRI APP44894 / UKRI 1281);  AI-Respire, ``AI for personalised respiratory health and pollution'' (EP/Y018680/1); ECO-AI, ``Enabling \ch{CO2} capture and storage using AI'' (EP/Y005732/1).

%\bibliographystyle{unsrtnat}
%\bibliography{references}

\end{document}